\documentclass[12pt,preprint]{aastex}

\shorttitle{Simulations on  a Galactic Gaseous Discs}
\shortauthors{Y\'a\~nez et al.}

\begin{document}

\title{Resonance Related Spiral Substructure in a Galactic Gaseous Disk}

\author{Miguel A. Y\'a\~nez,
Michael L. Norman}
\affil{Center for Astrophysics and Space Sciences, University of California at
    San Diego, La Jolla, CA 92093}
\email{myanez@ucsd.edu}
\email{mlnorman@ucsd.edu}

\author{Marco A. Martos}
\affil{Instituto de Astronom\'\i a, Universidad Nacional Aut\'onoma de 
M\'exico A. P.  70-264, M\'exico 04510, D. F.,  M\'exico}
\email{marco@astroscu.unam.mx}

\and

\author{John C. Hayes}
\affil{Lawrence Livermore National Laboratory, Livermore, CA 94550}
\email{jchayes@llnl.gov}

\begin{abstract}

We use high resolution (2048$^2$ zones) 2D hydrodynamic simulations to study
 the formation of spiral substructure in the gaseous disk  of a galaxy. 
The obtained gaseous response  is 
driven by a self-consistent non-axisymmetric potential  obtained from an 
imposed 
spiral mass distribution.
We
highlight the importance of ultraharmonic resonances in generating
these features. The temporal evolution of the system is followed with the 
parallel ZEUS-MP code, and we follow the steepening of perturbations 
induced by 
the spiral potential
until
large-scale shocks emerge. These shocks exhibit bifurcations that protrude 
from the 
gaseous arms and continue to steepen until  new shocks are 
formed.
When the contribution from the spiral potential relative 
to the axisymmetric background is increased from our default value, spurs 
protrude from the main arms after several revolutions of the gaseous disk.
Such spurs overlap
 on top of
the aforementioned  shocks. These results
support the hypothesis 
that a complicated gaseous response can coexist with an orderly spiral
potential term, in the sense that the underlying background potential can be 
smooth yet drive a gaseous response that is far more spatially complex.
\end{abstract}


\section{Introduction}

Young stars, H II regions and OB
associations delineate the structure that gives its name to spiral galaxies 
(Elmegreen 1981; Vall\'ee 2002, 2005). This implies a 
correlation between the 
spiral structure and
the process of star formation. The density wave theory attempts to explain the
large-scale structure of these galaxies in terms of a wave propagating in the 
disk of stars. Yet this stellar wave alone cannot directly explain the narrow 
spiral 
arms as delineated by the products of star formation.

Fujimoto (1968) first proposed that the dust lanes
observed on the concave side (inside corotation) of spiral arms might be the 
result of a Galactic 
shock. Large scale shocks propagating in the gaseous layer of the Galaxy could 
induce a sudden compression of the interstellar medium (ISM) and trigger the 
star formation process. 
Since the temporal scales involved in this process are large 
(of the order of the Galactic rotation period), an 
isothermal 
approximation of the ISM is a first step in modeling the gaseous response, 
thus rendering a highly 
compressible-supersonic medium in which a small amplitude disturbance, such as
that provided by an underlying spiral potential, may steepen into a shock 
wave. 

In a semi-analytical study of the gas flow in spiral density waves,
Roberts (1969) found nonlinear steady-state solutions containing shocks.
On the basis that these shocks were coincident with the imposed perturbing 
potential minima, he argued that the density jump could trigger star 
formation, and in
this way narrow bands of young stars could delineate the spiral arms.

Shu  et al. (1973) studied the gas flow in 
the context of the spiral density wave by adopting a two phase model for the 
ISM. In their study they carried out
a slightly nonlinear analysis of such flow and found that, at certain points 
in their 
solution, the amplitude  became infinite, and they called these 
positions ultraharmonic resonances. They argued that  additional 
prominent features could appear as a consequence of the ultraharmonic 
resonances, for instance: they found a secondary compression associated with
the first one.

However, these studies were carried out  assuming steady state flow
and tightly wound spirals. By removing the first constraint, Woodward (1975) 
showed how the convective steepening of the initial perturbation leads to 
large-scale shock formation (as a response to the driving potential) in a 
gaseous layer initially in 
circular orbit. 
He also found the effect of the first ultraharmonic resonance
as a secondary peak in the density field, but numerical viscosity in his 
calculations 
inhibited the formation of secondary shocks. 

In two papers, Contopoulos \&  
Grosb\o l (1986, 1988) removed the second constraint and
showed that nonlinear effects are not negligible
 in open spirals. Using numerical calculations, they 
demonstrated
that for open spirals the effect of the first 
ultraharmonic resonance (which they call the 4:1 resonance) is such that 
stellar orbits do not support the spiral perturbation beyond the position of 
this resonance 
and hence
the length  of the stellar spiral arms is limited by this position. In an 
analytical study of the effects of the ultraharmonic resonances,
 Artymowicz \& Lubow (1992) explained this result as a cumulative effect of 
higher-order resonances between  4:1 and the corotation radius. They further 
argued that in the vicinity of the former  the gas response is such that  it 
looks like a 4-arm structure with a smaller pitch angle  than that of 
the original 
two-arm pattern. 

It is clear that,  due to the nature of the problem, the highly nonlinear 
phenomena associated with the gaseous response to a spiral density wave are 
best studied via numerical simulations.
Results so obtained
add interarm features and spiral substructure on top of  large scale 
shocks. Patsis et al. (1997) found that  in open spirals  a 
bifurcation 
of the main spiral arms  takes place at the 4:1 resonance position. 
By analyzing numerical experiments that include self-gravity in the study 
of the 
effects of the ultraharmonic resonances on gaseous disks,
Chakrabarti et al. (2003, from here on CLS03) found secondary 
compressions, associated
with the first ultraharmonic resonance, in models
with a low Toomre parameter, Q, which were evolved only for a few revolutions 
of the disk. These compressions
eventually became branches.
 In high-Q models, evolved for several revolutions 
they found the appearance of leading structures which they identified with 
spurs. However, spurs and branches not related to resonances have also been 
obtained in numerical simulations (Dobbs \& Bonnell 2006; Wada \& Koda
2004). By taking into 
account 
frozen magnetic fields and self-gravity in 
their local arm simulations, 
Kim \& Ostriker (2002, 2006) showed that gaseous spurs form as a consequence 
of 
gravitational 
instability inside the spiral arms, a 
result 
confirmed with global 2D simulations by Shetty \& Ostriker (2006). In those
simulations spurs jut out of the spiral arms at regular intervals.

In this paper we carry out two dimensional, global, high resolution 
hydrodynamic simulations 
to further study the formation of spiral substructure in galactic gaseous 
disks.
Our approach differs from previous work in that we employ a self-consistent
model for the spiral stellar potential (in the orbital sense). This potential 
accounts for its own self-gravity and thus is no longer a local arm 
approximation to the driving term, making it  more appropriate for 
simulations of 
open 
galaxies.  This paper is organized as follows: in \S 2 we describe the 
potential we employ, in \S 3 we present the obtained gaseous response and the 
type of 
substructure related to it, in \S 4 we discuss the results and compare with 
previous work, and in \S 5 we present our conclusions.

\section{Modeling}
\subsection{The potential}

We employ a potential that consists of two parts. The first one is the 
axisymmetric background contribution, taken  from Allen 
\& Santill\'an (1991). This model assembles contributions from a bulge and
a flattened disk in the way proposed by Miyamoto and Nagai (1975), plus a 
massive spherical dark halo. This potential was chosen for its  
simplicity and easy mathematical manipulation. It renders a flat rotation
curve  of about 200 km s$^{-1}$ at a moderate radius and provides reasonable
values for Galactic parameters. 

The nonaxisymmetric contribution is more delicate.  
The usual approach
to model this term is via a logarithmic spiral potential. The rationale 
underlying this 
choice is two-fold: first, 
from the mathematical point of view it is tractable,
and second, no simpler form is known to represent this term. Yet, this choice 
has some limitations. It assumes  that only local effects of 
the potential can influence the gas dynamics; that is, it considers the 
potential minima to be in the geometrical center of the spiral arms. 
Such a potential is self-consistent only for small pitch angles 
 and does not consider its own 
self-gravity. 
To remove such
constraints and perform more adequate simulations in the presence of an open 
spiral, we choose a self-consistent potential 
(in the orbital sense) and with it performed our numerical experiments.
We employ 
a model derived from an imposed stellar spiral mass distribution
 given 
by Pichardo et al. (2003). 
In this paper self-consistency refers to the 
reinforcement of the spiral potential by the stellar orbits, i.e., the  orbits 
of the stars respond to the spiral potential in such a way as to reform the 
spiral locus of the initially imposed potential.

In order to obtain the potential given by an imposed open spiral mass 
distribution, Pichardo et al. (2003) selected a possible spiral locus for the 
morphology of the Galaxy. Then they place a series of oblate spheroids along 
this locus to fit the mass distribution of our Galaxy. In this way the total 
potential at a given point is found by adding the contributions of every 
spheroid at that same point.
The characteristics of the spheroids are as follows: the minor
axis of the spheroids is perpendicular to the Galactic plane and extends up to 
0.5 kpc. The major semiaxes  have a length of 1 kpc and lie in the Galactic 
plane. The locus on which 
they are placed fits the K band data of Drimmel (2000) with a pitch angle of 
15.5$^{\circ}$. The central density of the spheroids follows an exponential law
with a scale length of 2.5 kpc, approximately that of the near-infrared 
Galactic disk (Freudenreich 1998). 
The self-consistency study of the potential was carried out following the same 
procedure as Contopoulos and Grosb\o l (1986). To obtain the density response 
to
the spiral perturbation, Pichardo et al.(2003) assume that stars with circular 
orbits and rotating in the same sense as the spiral pattern are trapped around 
the corresponding central periodic orbit in the presence of the spiral 
perturbing term. They compute a series of central periodic orbits and find the 
density response along their extension. They match the density response maxima 
with the center of the assumed spiral arms 
(the imposed spiral locus) and whenever they match, self-consistency can be 
claimed.
In this way they showed that self-consistency of their models was a 
function of two parameters: the angular 
pattern speed ($\Omega_p$) and the ratio of 
mass contained in the arms to that contained in the disk, M$_A$/M$_D$.
The
best-fit values of these parameters to achieve self-consistency were 
$\Omega_p=$20 km s$^{-1}$ kpc$^{-1}$ and  
M$_A$/M$_D$=0.0175, respectively.

\subsection{Model Parameters}

Our numerical calculations compute the 2D hydrodynamic gaseous response to an 
external, fixed, spiral mass distribution in a 3D  galactic disk 
model. We performed our simulations in a frame that corotates with
the imposed spiral pattern.
The gaseous layer was initialized
in centrifugal equilibrium with
the axisymmetric potential and then the spiral term was turned on. The 
nonaxisymmetric term was modeled as a rigidly rotating potential with an 
angular
speed  ${\bf \Omega_p}$. We investigated spiral substructure formation by 
solving the
hydrodynamical equations in polar coordinates in a frame rotating with this 
prescribed 
pattern  speed.
 In this frame the hydrodynamical equations are

\begin{equation}
\frac{\partial \rho}{\partial t} +  \nabla \cdot (\rho \mbox{\bf{\em v'}}) =0
\label{continuity}
\end{equation}

\begin{equation}
\rho \left[ \frac{\partial \mbox{{\bf{\em v'}}}}{\partial t} +(\mbox{\bf{\em v'}} \cdot \nabla)\mbox{\bf{\em v'}} \right] 
=-\nabla p -\rho(\nabla \Phi_{as}+\nabla \Phi_s +2{\bf\Omega_p}\times \mbox{\bf{\em v'}} 
+{\bf \Omega_p}\times({\bf \Omega_p} \times r))
\label{euler}
\end{equation}

\begin{equation}
\rho\left[ \frac{\partial}{\partial t}\left( \frac{e}{\rho} \right) + {\mbox{\bf{{\em v'}}}}
\cdot \nabla \left( \frac{e}{\rho} \right) \right] + p\nabla \cdot {\mbox{\bf{\em v'}}}=0
\label{energy}
\end{equation}

In these equations $\rho$ is the gas density, {\bf {\em v'}} 
the velocity in the 
rotating frame, $p$ the 
pressure, $\Phi_{as}$ is the background axisymmetric potential, $\Phi_s$ the 
potential associated with the imposed spiral mass distribution, and $e$ the 
gaseous internal energy. 

An idealized ISM was assumed for these simulations in that we adopt an 
effectively isothermal
equation of state ($\gamma$ = 1.01) for the system. A constant sound speed, 
$c_s$, of 7 km s$^{-1}$ was 
used. We followed the evolution
of the gaseous response for several 
rotations; one rotation  is completed after $t_{orb}=2\pi /\Omega_0$ at a 
fiducial 
radius $R_0$. We present simulations that  evolve for 10 revolutions.

The effect of the first ultraharmonic resonance was studied by changing  
$\Omega_p$  so that the location of the resonance lay either inside or outside
the maximum radius of the spiral mass distribution (12 kpc). This was 
accomplished by selecting
$\Omega_p$ values of 20 and 10 km s$^{-1}$ kpc$^{-1}$, respectively. We also 
 explored the effects of a stronger spiral potential relative to the background
 axisymmetric contribution.
 Since the driving term
we use is obtained from   an imposed  mass distribution, we may control the 
relative force perturbation from the spiral potential by adjusting the mass 
contained in the spiral arms.
Using a constant pitch angle of 15.5$^{\circ}$ for the spiral locus, force 
perturbations of
3\% and 10\% are obtained from mass ratios for the spiral arms to the disk
 (M$_A$/M$_D$) of 0.0175, 
0.05,
respectively.

We follow the temporal evolution of the gaseous disk by solving the 
hydrodynamical equations~(\ref{continuity})-(\ref{energy}) in cylindrical 
polar 
coordinates (Z-R-$\phi$) with the ZEUS-MP
code (Hayes et al. 2006). ZEUS-MP differs from previous implementations of the 
ZEUS code in that it is designed for numerical calculations on massively 
parallel  computing platforms.
ZEUS-MP solves the equations of radiation magnetohydrodynamics in 1D, 2D or 3D
in cartesian, cylindrical or spherical  coordinates. The code
has been tested against an extensive bank of test problems 
documented in Hayes et al. (2006). For our simulations, we neglect  magnetic 
fields and radiation transport.

Our model uses
a computational mesh covering 2$\pi$ radians in azimuth and extending 
from 1 to 15 kpc in radius. Several boundary conditions were tried to ensure 
that wave reflections at the boundaries are not affecting our results. Inner 
inflow conditions prevent wave reflections from infalling gas and outer open 
conditions prevent gas from becoming stagnant. 
2D models were computed in cylindrical polar coordinates by modifying the code
so that symmetry was imposed along the leading (Z) coordinate and all 
acceleration terms along the Z axis in the momentum equation were zeroed.
We present numerical experiments
in a computational mesh with 2048$^2$ zones in R and $\phi$; we are  thus 
able to resolve the 
formation of substructure on  scales $\leq$  10 pc in the radial direction. 
The minimum and maximum resolution achieved on the computational grid are
 21 pc$^2$ and 293 pc$^2$ respectively. 

\section{RESULTS}

\begin{figure}
\epsscale{.9}
\plotone{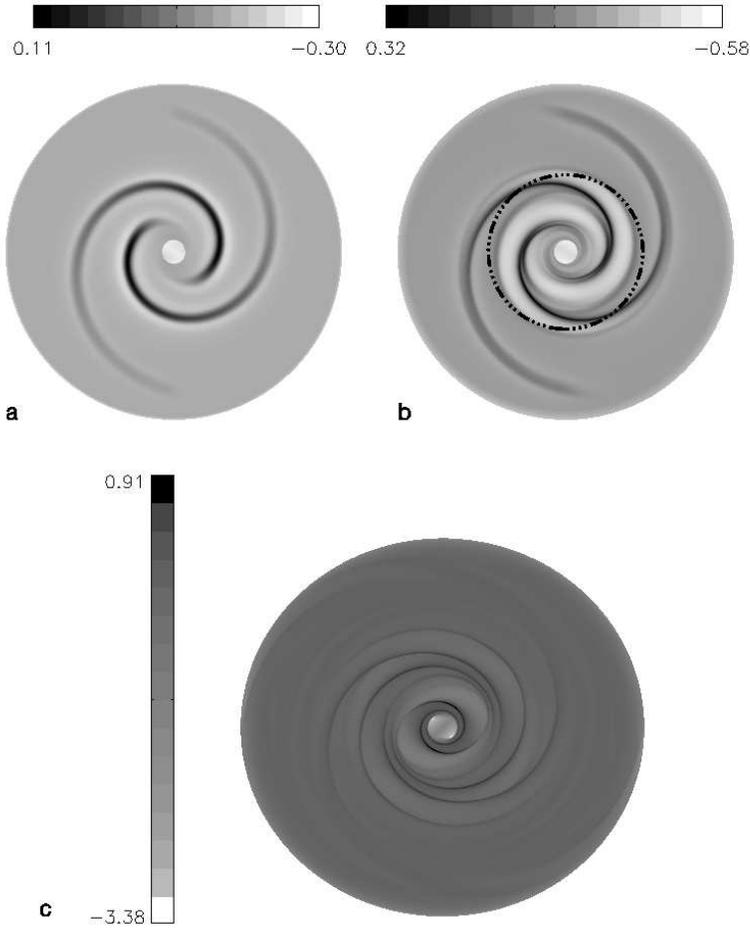}
\caption{Density snapshots for the case $\Omega_p$=20 km s$^{-1}$ kpc$^{-1}$. Surface density at (a) t/t$_{orb}$=0.1, (b) t/t$_{orb}$=0.26 the position of the first ultraharmonic resonance is marked by a dashed line, and (c) t/t$_{orb}$= 8.25. The density scale is shown in units of $\log  \Sigma/\Sigma_0$. The outer boundary for this simulation is located at 15 kpc.}
\label{om20}
\end{figure}

\begin{figure}
\epsscale{.9}
\plotone{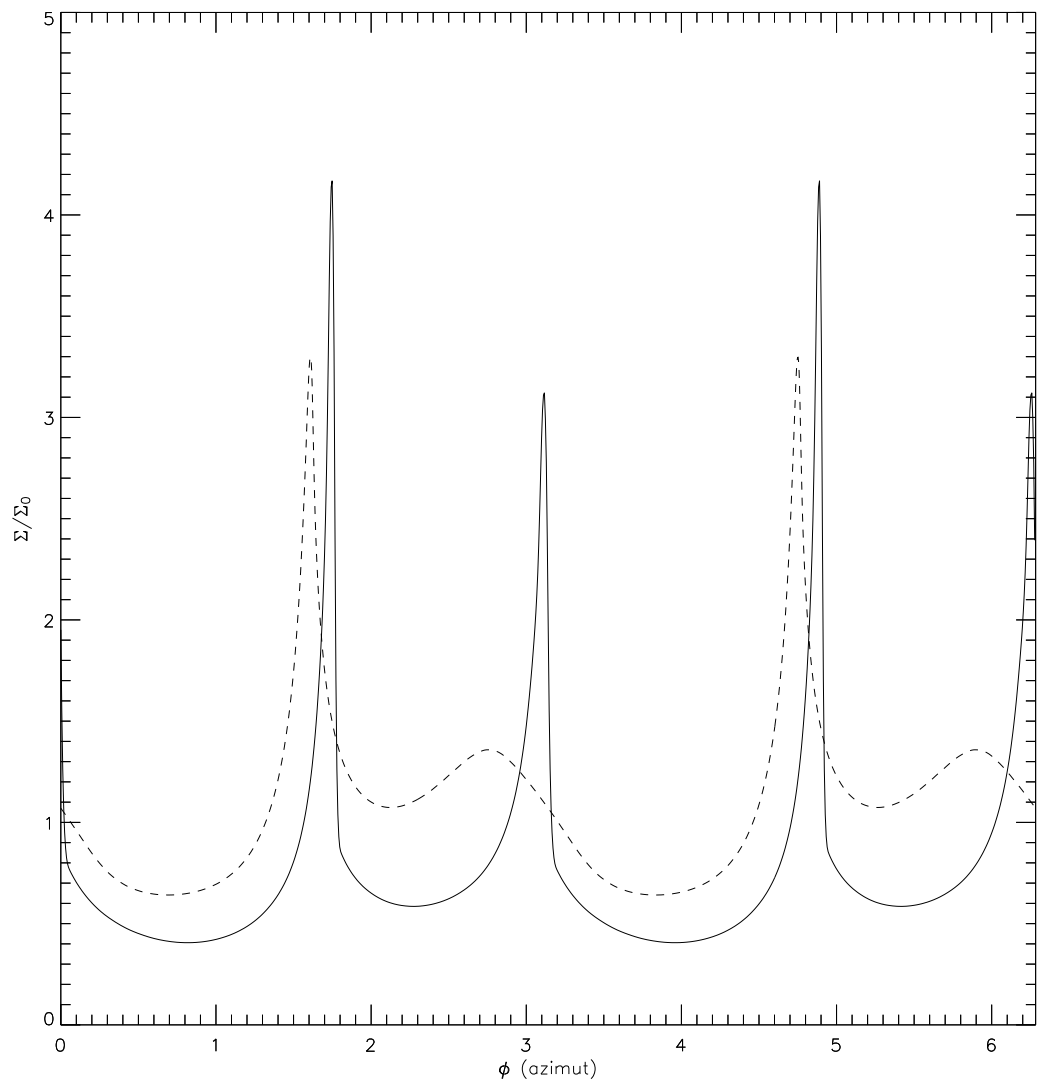}
\caption{Density cut at 6 kpc. We have plotted the ratio of  
density to initial density, to show the arm to interarm contrast 
 attained in this simulation. The dotted line represents density
contrasts at t/t$_{orb}=0.26$ and the continuous line at 
t/t$_{orb}=8.5$.}
\label{slice}
\end{figure}

In Figure~\ref{om20} we show snapshots of the density distribution  
corresponding
to the case  $\Omega_p=20$ km s$^{-1}$ kpc$^{-1}$.  We  observe that at 
 $t/t_{orb} =0.1$ the gas responds to the external potential by tracing the 
potential minima, clustering around them. At this time  the gas is adjusting 
to 
the potential and the density perturbation is steadily steepening in response 
to 
the spiral arms. Slightly
 later, at $ t/t_{orb}=0.26$, the recently formed gaseous arms begin to 
bifurcate. This bifurcation occurs in the vicinity of the first ultraharmonic 
resonance at a radius of 7 kpc, which is marked with a dashed line. Gas 
continues to accumulate around 
 the potential minima and the newly formed bifurcations until two pairs of 
large-scale shock waves are formed. We show the final steady-state 
configuration in the third snapshot of Figure~\ref{om20}. Two shock waves  
 trace the underlying 
spiral potential while the other two shock waves are the result of the 
steepening of
 the bifurcations. Similar results were presented by Martos et al. (2004) at 
lower resolution and with a previous version of the code. 
In Figure~\ref{slice} we
have plotted a density cut at a radius of 6 kpc, in order to 
illustrate the arm to interarm contrast at  evolution times corresponding to 
snapshots (b) and (c) of Figure~\ref{om20}. The arm to interarm contrast is a 
function of radius and time, but generally speaking we can state that the 
shocks associated with the imposed spiral mass distribution have a higher 
contrast and are more extended in the radial direction than the shocks 
associated with the bifurcations. The compressions, $\rho/\rho_0$, induced by 
these shocks 
have a peak around 8, in the inner regions, and 3 on average. The new shocks 
arising from  bifurcations of the first shocks induce lower compressions with
average values around 2. These shocks also have a lower pitch angle and their 
radial 
extent is limited by the first ultraharmonic resonance position. They can be
 found in the range of 6-7.5 kpc.

\clearpage

\clearpage

\subsection{Varying $\Omega_p$}

\begin{figure}
\epsscale{.9}
\plotone{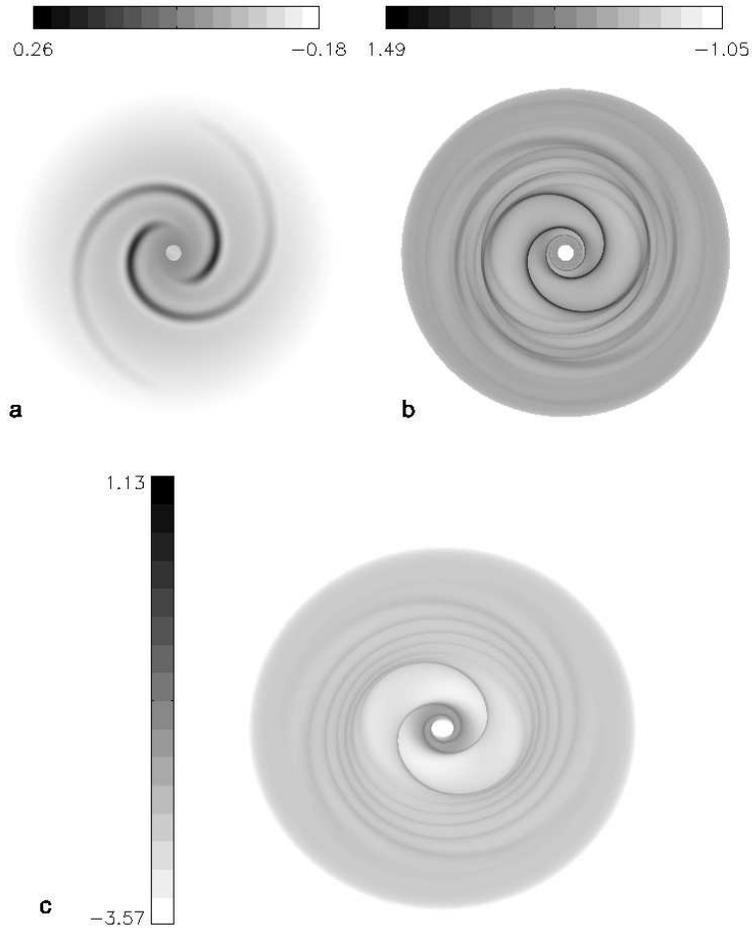}
\caption{Density distribution snapshots for the case $\Omega_p$=10 km s$^{-1}$ kpc$^{-1}$. Inner boundary is at 1 kpc and outer radius at 22 kpc. The position of the first ultraharmonic resonance is 14 kpc. Surface density at (a) t/t$_{orb}$=0.05, (b) t/t$_{orb}$=1.03, and (c) t/t$_{orb}$= 10. The density scale is shown in units of $\log  \Sigma/\Sigma_0$.}
\label{om10}
\end{figure}

In order to emphasize the importance of the first ultraharmonic resonance on
the bifurcation of the spiral arms, we have performed simulations where the 
position of this resonance has been shifted beyond the termination of the 
imposed spiral mass distribution. If we maintain fixed all other parameters 
and vary $\Omega_p$ from 20 to 10 km s$^{-1}$ kpc$^{-1}$ then the new position
of the first ultraharmonic resonance is 14 kpc. In 
Figure~\ref{om10} 
we show the density distributions for this case. 

\clearpage
\clearpage

The evolution of the system is similar to before:
the induced perturbations in the gas steepen into large-scale shock waves and 
accumulate
around the potential minima. 
In snapshot (b) we observe transient substructure emerging as waves propagate 
radially. After a few revolutions of the potential, this substructure is 
smoothed. In snapshot (c) only a two-arm gaseous response remains and the 
density peaks have been shifted downstream relative to the previous case. We 
also observe waves propagating from the tips of the arms radially outwards. 
These waves have been reflected in the corotation region and amplified, yet 
they do not add to the large-scale shocks and remain in an intermediate 
region between the end of the imposed potential and the corotation radius.

\subsection{Higher Forcing}

\begin{figure}
\epsscale{.9}
\plotone{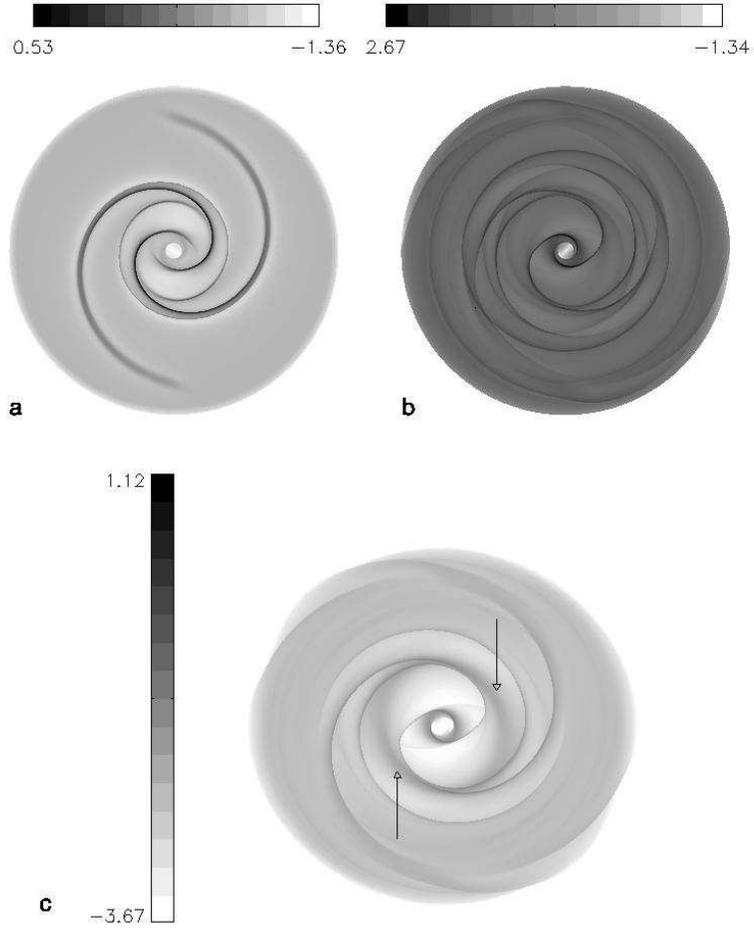}
\caption{Density contours snapshots for the case M$_A$/M$_D$=0.05. Surface density at (a) t/t$_{orb}$=0.26, (b) t/t$_{orb}$=1.55, and (c)
t/t$_{orb}$= 10. Spurs are delimited with solid lines, protruding from spiral arms. The density scale is shown in
units of $\log  \Sigma/\Sigma_0$.}
\label{highforcing}
\end{figure}

A higher forcing case corresponds to either a more massive or a more open 
spiral 
driving term. Since the potential model employed is derived from an imposed
spiral mass distribution, we increased the mass  in the spiral distribution to 
achieve a higher forcing term. In the case described here we adopted a mass
 ratio of 
the arms to the disk of M$_A$/M$_D$=0.05. With this parameter 
we have an average relative force perturbation of $~$ 10$\%$. The gaseous 
response presented in Figure~\ref{highforcing} is such that, initially, 
density enhancements around the potential 
minima are obtained. At t/t$_{orb}=0.26$ (snapshot a) branches begin to emerge 
near the 
first
 ultraharmonic 
position (at 7 kpc) and eventually interact with the already formed gaseous 
arms. Waves 
reflect from the inner Lindblad resonance (at 3 kpc) and the corotation radius 
(located at 11 kpc) and interact
with the existing density enhancements as they steepen. In this way, we 
obtain a temporary ragged appearance, illustrated in snapshot (b) at 
t/t$_{orb}$=1. The system, however, evolves to a quasi-steady state, and
this rich substructure is eventually sheared away. In this simulation the 
shock 
structure does not remain fixed; the shocks oscillate and so does the spiral 
substructure. We followed this simulation  to $t/t_{orb}$=10, and 
both oscillations and overall shock structure were preserved. In  
this sense  we say that system achieved a quasi-steady state. 

 An  interesting outcome of this 
simulation is the appearance after a few revolutions of substructure in the 
internal regions of the computational mesh. Short streams extend out from the 
main gaseous arms; these streams are small compared with the bifurcations and 
have a greater pitch angle. Because these characteristics match those given by 
Elmegreen (1981) to classify spurs, so we will refer to  them as such.
In 
snapshot (c) we present the final configuration of the system which displays
the previously noted  4-arm shock structure plus spurs at an evolution time 
of t/t$_{orb}$=10. The spurs are marked with draw-in arrows. These spurs 
protrude from the main arms into the already steepened branches, and as can be 
seen in this snapshot, they wind in an opposite sense to the general gas 
rotation. The spurs are symmetric with respect to the main arms and originate
 close to inner Lindblad resonance.

\clearpage
\clearpage

\section{Discussion}

Analytical work on the gaseous disk response to an external driving spiral 
potential
by Shu et al. (1973) and Artymowicz \& Lubow (1992) highlighted
the importance of the first ultraharmonic resonance as the main generator of
spiral substructure. Those  studies concluded that in a slightly nonlinear 
regime a second compression emerged in a gaseous disk that could account for
observed spiral substructure. But a more complete study is necessary to 
identify 
 the effect of resonances in a fully nonlinear regime. Due to the 
nature of the equations governing the evolution of such a disk, this task is 
best accomplished using numerical simulations.  

Numerical work on this subject by Patsis et al. (1997) and
CLS03
revealed that substructure arising in gaseous disks can be related to 
ultraharmonic 
resonances.

The gaseous response we have presented resembles that of previous numerical 
work: the initial response tends to associate with the underlying potential,
and once gaseous arms have formed, branches begin to protrude from them.
 But in our simulations these branches 
continue to steepen,  leading to the formation of new shock waves, with 
lower radial extension, pitch angle and Mach number, but large-scale shock 
waves 
nevertheless. The role of the first ultraharmonic resonance is such that,
if absent from the region of influence of the spiral term, no such 
additional shocks are formed.

We  also 
performed simulations with high forcing and  obtained additional 
substructure:  gaseous spurs. These spurs protrude from the arms after
several revolutions and are not a transient feature.
 We  wish to note  that these results were 
obtained without 
self-gravity in the gaseous disk and that spurs and bifurcations are obtained 
in the same long-timescale simulation.
 Such spurs  wind in the opposite sense  to the arms 
rotation  and are restricted to an area close to the inner Lindblad resonance.
CLS03 also obtained gaseous spurs employing a nonlinear perturbing potential 
with high forcing in a disk very stable against 
axisymmetric
 perturbations, but bifurcations did not appeared in those simulations.
 In our case we use a spiral term that ensures self-consistency and
employ the maximum value for forcing (5\%, a similar value as that used by 
CLS03 to obtain spurs) that still renders a self-consistent
driving term according to Pichardo et al. (2003). Our driving term and our 
algorithms to calculate the gaseous response have proven robust enough as to 
allow the development of branches and spurs in the same run. The high numerical
resolution and consequent low numerical dissipation of our simulation may also
have been important in obtaining this result.
CLS03 include self-gravity in 
the gaseous layer and the spurs they obtain are more pronounced. Although we 
demonstrated that these features can emerge in the 
absence of self-gravity in the gaseous layer, we are well aware that
self-gravity in the gas 
 will enhance the  features we have found giving rise to higher arm-interam
density contrasts.

It is worth remarking that the simulations presented here can be 
evolved for several revolutions of the driving term (more than t/t$_{orb}=10$)
and eventually achieve a steady state. Yet the isothermal condition we assumed
can only reproduce the formation of substructure, but not its subsequent 
fragmentation. To reproduce the latter
 we need to adopt a more realistic treatment of the
ISM, with special attention to its thermodynamic properties. Another line
of improvement is suggested by previous work  showing that the inclusion 
of frozen magnetic fields  leads to the formation of additional substructure. 
We
defer a discussion of the role of magnetic fields to a future paper.

\section{Conclusions}

A fully nonlinear treatment of the propagation of spiral density waves in a 
gaseous disk is tractable using  numerical simulations.  
 By considering an open spiral, in the absence of self-gravity in the gaseous 
layer, we are able to 
obtain rich substructure associated with an external spiral potential. Our work
differs from other published results in that we employ a self-consistent
driving term that considers its own gravity and removes the local arm 
approximation. We have
presented experiments showing the formation of a four-arm structure in 
response to a two-arm driving pattern. Initially the gas accumulates in the 
potential minima. After some time the gas responds to resonances and the main
arms bifurcate. These features have a considerable azimuthal extension and 
continue to 
steepen with  time. Eventually a four-arm shock structure  emerges, in the 
gaseous layer that 
coexist with 
the underlying two-arm potential, made up by the stars. This result appears to 
  agree with
observational data published by Drimmel (2000) where he concludes that, using
optical tracers, the spiral structure in our Galaxy is best fitted by a 
four-arm 
 structure, 
while in the infrared a two-arm structure dominates.

If we place the first ultraharmonic resonance outside the region of influence 
of the spiral (in our case by changing the value of $\Omega_p$) the main arms 
do not bifurcate and large-scale shocks are the only induced response in the 
gaseous layer. In this way we show that if the first ultraharmonic resonance is
 placed outside the region of influence of the perturbing term, no spiral 
substructure appears, thus emphasizing the connection between this resonance
and the bifurcations of the gaseous arms.

If we combine the effects of the resonances with a large 
forcing amplitude we obtain, on top of this four-arm structure, 
spurs protruding from the main arms in a region between the inner Lindblad 
resonance and the first ultraharmonic resonance. Overlapping of nonlinear 
effects take place in our long-timescale global simulations, showing that 
an orderly spiral density wave potential can produce a gaseous
response that is strongly disordered.

The inclusion of additional processes such as magnetic fields, self-gravity 
for 
the gaseous layer, and
thermal processes in the ISM may lead to the appearance of additional 
substructure and 
their 
consequent fragmentation into bound condensations. Numerical experiments 
addressing these topics are needed in order to improve our understanding of 
the 
global ISM in galaxies and its relation to large-scale processes.

\acknowledgments
M. Y\'a\~nez and M. Martos want to thank the UCMEXUS-CONACYT fellowship 
program for
economic support. M. Martos thanks the CASS-UCSD for their hospitality during 
a sabbatical visit which helped pursue this project.
The numerical simulations were carried out on the DataStar
system at the San Diego Supercomputer Center with LRAC allocation MCA98N020.

\clearpage

\end{document}